\documentclass[iicol,referee,pdflatex,sn-aps]{sn-jnl}
\usepackage{lineno}
\usepackage[numbers,sort&compress]{natbib}
\usepackage{amsmath,amsmath,amsfonts}
\usepackage{mathrsfs}
\usepackage{graphicx}
\usepackage{ulem}
\usepackage{subfigure}%
\usepackage{array}
\usepackage{diagbox}
\usepackage{color}
\usepackage{multirow,cuted}
\usepackage[title]{appendix}
\allowdisplaybreaks[4]
\date{\today}

\begin{document}
	\title{Gravitational orbital Hall effect of vortex light in Lense-Thirring metric}
	\author[1]{\fnm{Wei-Si} \sur{Qiu}}
	\email{qiuws@mail2.sysu.edu.cn}
	\author*[1]{\fnm{Dan-Dan} \sur{Lian}}
	\email{liandd@mail.sysu.edu.cn}
	\author[1]{\fnm{Peng-Ming} \sur{Zhang}}
	\email{zhangpm5@mail.sysu.edu.cn}
	\affil[1]{School of Physics and Astronomy, Sun Yat-sen University, 519082 Zhuhai, China}

\abstract{
Vortex light, characterized by an intrinsic orbital angular momentum aligned with its propagation direction, is described through vortex electromagnetic waves. Similar to the gravitational spin Hall effect (SHE), vortex light is expected to exhibit intrinsic orbital angular momentum dependent trajectories and deviations from the null geodesic plane when propagating through a gravitational field, a phenomenon termed the gravitational orbital Hall effect (OHE). In this work, we model the vortex light as vortex Laguerre-Gaussian electromagnetic wave packets and analyze its motion by solving covariant Maxwell equations within the Lense-Thirring metric. Our findings reveal that the trajectory of vortex light with an intrinsic orbital angular momentum deviates from the null geodesic in two ways. It deviates both perpendicular to, and within, the null geodesic plane. This behavior contrasts with the gravitational SHE, where spin-polarized light primarily deviates perpendicular to the null geodesic plane. Moreover, the relationship between the deviation and intrinsic orbital angular momentum differs significantly from that between the deviation and spin. These results suggest a unique interaction between intrinsic orbital angular momentum and gravity, distinct from the spin-gravity coupling, indicating that the gravitational OHE of light might not be precisely predicted by merely substituting spin with intrinsic orbital angular momentum in the gravitational SHE of light.
}
\maketitle
\section{Introduction}\label{sec.Intro}

The spin Hall effect (SHE), characterized by spin-dependent trajectory deviations of particles due to spin-orbit coupling in specific media or external fields, is a well-documented phenomenon in light propagation through gradient-index media and various optical materials. This effect has been thoroughly discussed, revealing both its detectability and foundational mechanisms \cite{bliokh2006nonlinear,hosten2008observation,bliokh2008geometrodynamics,bliokh2009spin,zhou2013photonic,liu2015photonic,bliokh2015spin,ling2023photonic}. Recent advancements have investigated the ability to control the SHE through crystal birefringence, highlighting its utility in the precise manipulation of micro-particles \cite{fu2019spin} and the behaviors in optical fibers with curved step-index profiles \cite{mieling2023polarization}. Authoritative reviews have effectively summarized the core concepts and varied applications of the optical SHE, significantly enhancing our comprehension of this field \cite{bliokh2015spin,ling2017recent,liu2022photonic,sheng2023photonic}.

In gravitational fields, particles also exhibit trajectories and deviations that depend on their spin, due to gravitational spin-orbit coupling, an effect known as the gravitational SHE. This phenomenon was initially identified by Papapetrou and Corinaldesi using the Mathisson-Papapetrou-Dixon (MPD) equations \cite{papapetrou1951spinning,corinaldesi1951spinning,dixon1970dynamics,Dixon:1970zz,1974RSPTA.277...59D,dixon2015new}. Subsequent studies have extensively investigated the gravitational SHE across various metrics employing these equations \cite{Carmeli:1976mq,Plyatsko:2011gf,Hackmann:2014tga,Antoniou:2019lit,duval2019gravitational,Zhang:2020qew}. However, deriving unique trajectories from the MPD equations requires the application of supplementary conditions to the angular momentum $S^{\mu\nu}$. Different supplementary conditions result in differing trajectories \cite{Costa:2014nta}.

Besides the MPD equations, researchers have employed other methods to study the gravitational SHE. These methods include applying the Wentzel-Kramers-Brillouin approximation to Maxwell's equations \cite{PhysRevD.86.024010,oancea2020gravitational,PhysRevD.104.025006,PhysRevD.105.104061,Andersson_2023,oancea2024weyl}, utilizing the energy-momentum tensor \cite{lian2022birefringence,lian2023motion}, and using Foldy-Wouthuysen transformation to derive Hamiltonians \cite{Obukhov:2000ih,Silenko:2004ad,PhysRevD.75.084035,Obukhov:2009qs,Obukhov:2013zca,PhysRevD.109.044060}. These studies not only deepen our understanding of the interactions between matter and gravity but also provide numerous predictions that can be used to test gravitational theories.

Beyond possessing spin, a single particle can also exist in a vortex state, carrying an intrinsic orbital angular momentum  aligned with its average momentum. The vortex state, can be described by the wave function $\psi \propto e^{i \ell \phi} $, with $\ell \hbar$ as its intrinsic orbital angular momentum \cite{landau2013quantum,cohen1977diu}. Here, $\phi$ is the azimuthal angle. These particles, when in such vortex states, are referred to as "vortex" particles, highlighting their intrinsic orbital angular momentum. In 1992, vortex states, characterized by Laguerre-Gaussian modes, were first discoveried by Allen et al. \cite{beijersbergen1993astigmatic}. Subsequently, in 1996, these modes were experimentally generated at millimeter-wave frequencies using a spiral phase plate \cite{TURNBULL1996183}. In the same year, the intrinsic orbital angular momentum of the Laguerre-Gaussian modes was experimentally quantified through the trapping of absorbent particles \cite{PhysRevA.54.1593}. Since then, vortex light has been generated in a variety of optical experiments and used in industrial laser production \cite{Bazhenov2003LaserBW,oemrawsingh2004production,gibson2004free,Lin2013NanostructuredHF,PhysRevLett.112.235001,Davis2015AnalysisOA,bliokh2015spin,RosalesGuzmn2017HowTS,bliokh2017theory}, demonstrating their applicability in numerous areas of fundamental physics. These applications include optical trapping of particles \cite{gahagan1996optical,garces2003observation}, quantum information and communications \cite{mair2001entanglement,leach2010quantum}, astronomy and astrophysics \cite{foo2005optical,harwit2003photon}, and optical solitons \cite{kruglov1985spiral,swartzlander1992optical}.

Similar to the SHE, the dynamics of particles are also affected by intrinsic orbital angular momentum when passing through materials, a phenomenon termed the orbital Hall effect (OHE). The OHE has been explored in light moving along spatial curves \cite{lai2018electromagnetic} and observed in electrons in titanium metals \cite{choi2023observation}. These findings lead to the expectation that vortex light, possessing intrinsic orbital angular momentum, will similarly exhibit altered dynamics in gravitational fields, an effect we term the gravitational OHE. However, the MPD equations and WKB approximation, typically used for the gravitational SHE analysis, are less effective for the gravitational OHE due to their inadequate description of vortex states and inability to distinguish between spin and intrinsic orbital angular momentum.

In this study, we aim to explore the gravitational OHE for vortex light and its potential distinctions from the gravitational SHE. Our investigation adopts a comprehensive methodology consisting of three primary steps:
\begin{enumerate}
	\item  We model the vortex light as Laguerre-Gaussian electromagnetic wave packets to capture their intrinsic orbital angular momentum.
	\item  We investigate the dynamics of vortex wave packets by solving the covariant Maxwell equations within the Lense-Thirring metric, enabling us to understand the influence of gravity on vortex light motion.
	\item  We assess the motion of the vortex light by analyzing the center of its energy-momentum tensor, which provides insights into the trajectory alterations induced by varying intrinsic orbital angular momentum.
\end{enumerate}
This methodology has been effectively utilized in previous studies to analyze the gravitational SHE of light within the Schwarzschild metric \cite{lian2022birefringence} and the gravitational OHE of scalar vortex particles in a stellar gravitational field \cite{lian2023motion}.

The structure of the paper is organized as follows: Section \ref{sec.dynam} introduces the vortex light as a Laguerre-Gaussian electromagnetic wave packet and explores its evolution by numerically solving the Maxwell equations within the Lense-Thirring metric. In Section \ref{sec.OHE}, we examine the gravitational OHE for vortex light in this metric, using the center of the energy-momentum tensor. This section also compares the gravitational OHE with the gravitational SHE and seek out the differences between these two phenomena. Section \ref{sec.dis} offers further discussion on our findings. Throughout this paper, we adopt the metric signature $(-,+,+,+)$ and normalize units such that $c=\hbar=1$.

\section{Dynamics of vortex light in the Lense-Thirring metric}\label{sec.dynam}

The dynamics of vortex light in vacuum, represented as electromagnetic wave packets, are governed by source-free Maxwell equations. Within a curved spacetime, these equations are expressed as
\begin{equation}\label{maxwell}
\nabla_\mu F^{\mu\nu}=0,
\end{equation}
where $F^{\mu\nu}=\nabla^\mu A^\nu (\vec{x},t)-\nabla^\nu A^\mu (\vec{x},t)$ denotes the electromagnetic tensor. When considering the motion of light outside a gravitational source, where the Ricci tensor $R_{\alpha\beta}$ vanishes, employing the covariant Lorenz gauge condition, $\nabla_\mu A^\mu (\vec{x},t)=0$, simplifies Eq. \eqref{maxwell} to
\begin{equation}\label{eom}
\nabla^\mu \nabla_\mu A^\nu (\vec{x},t)=0.
\end{equation}

In the weak field approximation, the curved spacetime around a slowly rotating star is well-described by the Lense-Thirring metric \cite{adler2015three,wald2010general}. In Cartesian coordinates, the line element $\text{d}s$ is defined as:
\begin{align}\label{metric}
\mathrm{d} s^{2}&=-\left(1-\frac{2GM}{r}\right)\mathrm{d} t^2+\left(1+\frac{2GM}{r}\right)\mathrm{d} x^2\nonumber\\
&+\left(1+\frac{2GM}{r}\right)(\mathrm{d} y^2+\mathrm{d} z^2)\nonumber\\
&-\frac{4GMa}{r^3}(x\mathrm{d}y \mathrm{d}t-y\mathrm{d}x \mathrm{d}t),
\end{align}
where $M$ represents the mass of the gravitational source, $r=\sqrt{x^2+y^2+z^2}$ the distance from the source's center, and $a=|\vec{J}|/M$ the angular momentum per unit mass, with the source's angular momentum $\vec{J}$ aligned along the $z$-axis. The Lense-Thirring metric is thus approximated by:
\begin{equation}
g_{\mu\nu}\simeq \eta_{\mu\nu} + h_{\mu\nu},
\end{equation}
with $\eta_{\mu\nu}=\text{diag}(-1,1,1,1)$ and the perturbation $h_{\mu\nu}$ as
\begin{equation}\label{pert-h}
	h_{\mu\nu}=\begin{pmatrix}
		\frac{2GM}{r}&\frac{2GMay}{r^3}&-\frac{2GMax}{r^3}&0\\
		\frac{2GMay}{r^3}&\frac{2GM}{r}&0&0\\
		-\frac{2GMax}{r^3}&0&\frac{2GM}{r}&0\\
		0&0&0&\frac{2GM}{r}\\
	\end{pmatrix}
\end{equation}

Within the framework of the weak field approximation, where $GM/r\ll 1$, it is practical to consider the gravitational impact on the electromagnetic field $A^\nu (\vec{x},t)$ as a perturbative effect:
\begin{equation}
A^\nu (\vec{x},t)=\bar{A}^\nu (\vec{x},t)+\widetilde{A}^\nu (\vec{x},t),
\end{equation}
with $\bar{A}^\nu (\vec{x},t)$ representing the zeroth-order term in $GM/r$, satisfying the equation $\eta^{\rho\mu}\partial_\rho \partial_\mu \bar{A}^\nu (\vec{x},t)=0$. According to Eq. \eqref{eom}, the perturbation $\widetilde{A}^\nu (\vec{x},t)$, being first order in $GM/r$, is approximately described by
\begin{align}\label{eom-first}
	\partial_\lambda\partial^\lambda\widetilde{A}^\nu (\vec{x},t) &\simeq h^{\lambda\alpha}\partial_\lambda\partial_\alpha\bar{A}^\nu (\vec{x},t)\nonumber\\
	&+\eta^{\lambda\alpha}\widetilde{\Gamma}^\beta_{\lambda\alpha}\partial_\beta\bar{A}^\nu (\vec{x},t)\nonumber\\
	&-\partial^\alpha\widetilde{\Gamma}^\nu_{\alpha\beta}\bar{A}^\beta (\vec{x},t)\nonumber\\
	&-2\widetilde{\Gamma}^\nu_{\alpha\beta}\partial^\alpha\bar{A}^\beta (\vec{x},t),
\end{align}
where $\widetilde{\Gamma}_{\lambda\alpha}^{\beta}=\frac{1}{2} \eta^{\beta \gamma}(\partial_{\alpha} h_{\lambda \gamma}+\partial_{\lambda} h_{\alpha \gamma}-\partial_{\gamma} h_{\lambda \alpha})$ denotes the first-order terms in $GM/r$ of the affine connection.

\subsection{Physical system and initial states }

\begin{figure}
\centering
\includegraphics[width=0.4\textwidth]{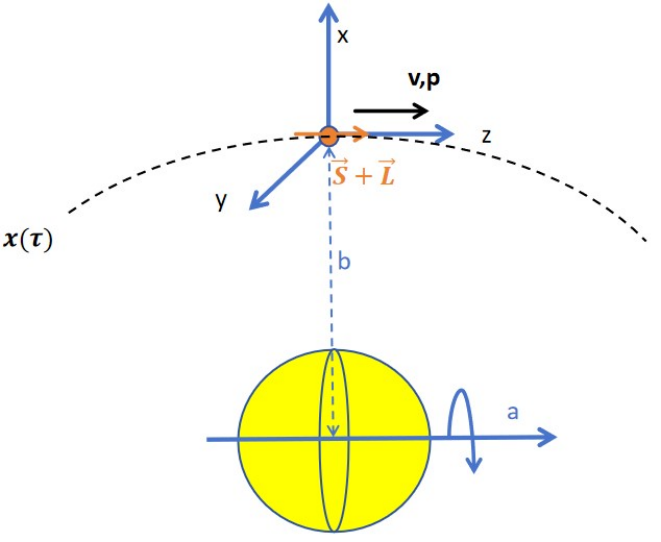}
\caption{Configurations of the physical system. The yellow sphere depicts the rotating gravitational source situated at $(-b,0,0)$. A vortex photon, symbolized by the orange sphere, initially positioned at the coordinate origin, starts its free fall along the $z$-axis. The dashed line $x(\tau)$ symbolizes the vortex photon's trajectory.}\label{models}
\end{figure}

The physical system under investigation, as depicted in Fig. \ref{models}, consists of a gravitational source located at $(-b,0,0)$, rotating around the $z$-axis. A vortex electromagnetic wave packet, initially positioned at the coordinate origin, commences its free fall with momentum $\vec{p}$ directed along the $z$-axis. Both its spin $\vec{S}$ and intrinsic orbital angular momentum $\vec{L}$ are aligned with the $z$-axis.

Our calculations reveal that when the wave packet's momentum is perpendicular to the gravitational source's rotational axis (the $z$-axis in Fig. \ref{models}), the trajectory of the wave packet is negligibly affected by the source's rotation. This observation leads us to primarily focus on the case where the momentum is initially aligned parallel to the rotational axis, allowing for a targeted investigation into the dynamics of the wave packet under these specific conditions.

In the absence of gravity within Minkowski spacetime, a vortex photon can be characterized by a vortex Laguerre-Gaussian electromagnetic wave packet \cite{liu2023threshold}. The expression for this wave packet in momentum space is given by
\begin{align}\label{LG-wave}
	\bar{A}_{f}^i(\vec{k},t) &= \frac{\sqrt{2} N (\sigma_{\perp} k_{\perp})^{\ell}}{\omega^{3/2}} \epsilon^i \exp \left(i \ell \phi_{k} - i\omega t\right)\nonumber\\
	&\times \exp \left(- \frac{(k_{z} - p_{z})^{2} \sigma_{z}^{2}}{2} -\frac{k_{\perp}^{2} \sigma_{\perp}^{2}}{2} \right),
\end{align}
where $\vec{\epsilon}=\{\epsilon^i\} = (k_{z}^{2} + k_{y}^{2} - i \sigma k_{x} k_{y}, i \sigma k_{z}^{2} + i \sigma k_{x}^{2} - k_{x} k_{y}, -i \sigma k_{y} k_{z} - k_{x} k_{z})^\text{T}$ with $\sigma=0,\pm 1$ represents the polarization vector, satisfying $\vec{k}\cdot \vec{\varepsilon}=0$. The wave vector's transverse component is denoted as $k_{\perp} = \sqrt{k_x^2 + k_y^2}$. The angular frequency is defined as $\omega = |\vec{k}| =\sqrt{k_x^2 + k_y^2 + k_z^2}$, and the azimuthal angle is given by $\phi_k = \arctan(k_y/k_x)$. The parameter $p_z$ refers to the average initial momentum of the wave packet along the $z$-axis, and $N$ is the normalization factor. This wave function represents a particular Laguerre-Gaussian mode in momentum space. As outlined in Section \ref{sec.Intro}, it is feasible to generate this wave packet experimentally, allowing for the potential observation of its gravitational OHE in principle.

In coordinate space, the wave packet can be represented through the inverse Fourier Transform as:
\begin{equation}\label{LG-coor}
\bar{A}^i(\vec{x},t)=\int \bar{A}^i_f(\vec{k},t)\exp (i\vec{k}\cdot \vec{x})\text{d}^3k.
\end{equation}
This wave packet adheres to the Coulomb gauge, $\partial_i\bar{A}^i(\vec{x},t)=0$ with $\bar{A}^0(\vec{x},t)=0$,  due to the orthogonal relation between the momentum and polarization vector $\vec{k}\cdot \vec{\varepsilon}=0$. Under these conditions, the source-free Maxwell equations in Minkowski spacetime simplify to $\partial_\mu\partial^\mu \bar{A}^\nu (\vec{x},t)=0$. Given that the wave packet's angular frequency in momentum space, is given by $\omega=|\vec{k}|$, it confirms that $\bar{A}^i(\vec{x},t)$ satisfies these source-free Maxwell equations in Minkowski spacetime.

For an electromagnetic wave packet, its spin and orbital angular momentum can be represented as 
\begin{align}
	\vec{S}=\int \vec{E}\times \vec{A}(\vec{x},t)\text{d}^3 x
\end{align}
and
\begin{align}
	\vec{L}= \int \vec{x}\times E^i \vec{\nabla}A^i (\vec{x},t)\text{d}^3 x,
\end{align}
where $\vec{E}=\{E^i\}=\{-\partial_t A^i(\vec{x},t)-\nabla_i A^0(\vec{x},t)\}$ denotes the electric field of light in coordinate space. Through straightforward calculations, we confirm that the wave packet $\bar{A}^i (\vec{x},t)$ carries a spin of $\sigma$ and an intrinsic orbital angular momentum of $\ell$ along its average momentum. Furthermore, we establish that the wave packet's initial momentum is directed along the $z$-axis, with its initial position at the coordinate origin. Therefore, this wave packet conforms to the configurations of the physical system under consideration in this study.

The radius of a realistic electromagnetic wave packet is expected to be significantly smaller than its distance, $b$, from the center of the gravitational source. For the wave packet described in Eq. \eqref{LG-wave}, its radius, $R$, can be approximated by $R \sim \sigma_\perp \sim \sigma_z$. Hence, we establish the following relationship:
\begin{equation}
	\sigma_\perp \sim \sigma_z \ll b.
\end{equation}
This indicates that the wave packet is confined to a significantly small volume in three-dimensional space, with its radius being much smaller than the distance to the gravitational source. Consequently, the gravitational field within the wave packet can be approximated as follows:
\begin{align}\label{h-beg}
&h_{00} \simeq \frac{2GM}{b}\left(1-\frac{x}{b}\right), \\
&h_{01} \simeq \frac{2GMay}{b^3},\quad h_{03}=0,\\
&h_{02} \simeq -\frac{2GMa}{b^2}\left(1-\frac{2x}{b}\right),\\
&h_{ij} \simeq \frac{2GM}{b}\left(1-\frac{x}{b}\right)\delta_{ij},\label{h-end}
\end{align}
where only terms of the first order in $x^i/b$ have been considered. Under these approximations, Eq. \eqref{eom-first} can be significantly simplified, as detailed in Appendix.

However, as shown in Appendix, the equation of motion for $\widetilde{A}^\nu (\vec{x},t)$ remains a set of second-order inhomogeneous partial differential equations, posing significant challenges for direct solution. Additionally, the wave packet of the vortex light is articulated in momentum space, necessitating the transformation of Eq. \eqref{eom-first} into momentum space through the following Fourier Transform:
\begin{equation}
	\mathscr{F}(f(\vec{x},t))=\frac{1}{(2\pi)^3}\int f(\vec{x},t) \exp(-i\vec{k}\cdot x)\text{d}^3 x.
\end{equation}
By applying this Fourier transform to Eq. \eqref{eom-first}, with the wave packet $\bar{A}_f^i (\vec{k},t)$ described in Eq. \eqref{LG-wave}, we reformulate these equations as:
\begin{equation}\label{eom-k}
    \left(\partial_{t}^{2}+\omega^{2}\right) \tilde{A}_{f}^{\rho}(\vec{k},t)=a_1^{\rho}(\vec{k}) e^{-i \omega t} +a_2^{\rho}(\vec{k}) e^{-i \omega t} t,
\end{equation}
where $\widetilde{A}^\rho_f (\vec{k},t)=\mathscr{F}(\widetilde{A}^\rho (\vec{x},t))$ represents the perturbation term in momentum space. The coefficient $a_1^{\rho}(\vec{k})$ encompasses all factors associated with terms that include the time-dependent factor $\exp(-i\omega t)$. Similarly, the coefficient $a_2^{\rho}(\vec{k})$ encapsulates all factors related to terms that incorporate the time-dependent expression $t\exp(-i\omega t)$. Detailed expressions of these equations are available in Appendix \ref{sec.app1}.

Given the initial conditions:
\begin{equation}\label{initial conditions}
    \left.\tilde{A}_{f}^{\rho}(\vec{k},t)\right|_{t=0}=0 \quad \text{and} \quad \left.\partial_{t} \tilde{A}_{f}^{\rho}(\vec{k},t)\right|_{t=0}=0,
\end{equation}
Eq. \eqref{eom-k} can be analytically solved. The perturbation term $\widetilde{A}^\rho_f (\vec{k},t)$ is expressed as:
\begin{align}\label{general solution}
    \tilde{A}_{f}^{\rho}(\vec{k},t) &=  \frac{a_1^{\rho}(\vec{k})(1-e^{2 i {\omega} t}+2 i {\omega} t) e^{-i {\omega} t}}{4 {\omega} ^{2}}\nonumber\\
	& + \frac{a_2^{\rho}(\vec{k})(2 {\omega} t-i+2 i {\omega}^{2} t^{2}+i e^{2 i {\omega} t}) e^{-i {\omega} t}}{8 {\omega}^{3}}.
\end{align}
Accordingly, the wave packet of the free-falling vortex light in coordinate space is determined through the inverse Fourier Transform: 
\begin{equation}
    A^{\rho}(\vec{x},t)=\int {A}^{\rho}_f(\vec{k},t) \exp (i \vec{k} \cdot \vec{x}) \, \text{d}^{3}k, 
\end{equation}
where $A^{\rho}_f(\vec{k},t)=\bar{A}^{\rho}_f(\vec{k},t)+\widetilde{A}^{\rho}_f(\vec{k},t)$.

Due to the complexity of the wave packet $A^\rho_f(\vec{k},t)$, analytically calculating the integration in the aforementioned equation is nearly infeasible. This integration is essentially an inverse Fourier Transform, which lends itself to numerical methods. Therefore, we first discretize the wave packet $A^\rho_f(\vec{k},t)$ in momentum space. Subsequently, we employ the "InverseFourier" function, a numerical tool for inverse Fourier Transform provided by Wolfram Mathematica (developed by Wolfram Research), to carry out this integration. This approach allows us to numerically derive the wave packet's distribution, $A^{\rho}(\vec{x},t)$, in coordinate space. Following these numerical computations, the energy-momentum tensor and the center of the vortex light can be straightforwardly calculated.

\section{Gravitational orbital Hall effect for vortex light}\label{sec.OHE}

The dynamics of vortex light are described through electromagnetic wave packets, which preclude their simplification to point-like particles. Consequently, traditional trajectories applicable to point-like particles might not be suitable for vortex light. In this study, we characterize the motion of vortex light by employing energy-momentum tensor, thereby defining their trajectories via the energy-momentum tensor:
\begin{equation}\label{center}
	\langle x^i\rangle^{\mu\nu}=\frac{\int \sqrt{-g}x^i T^{\mu\nu}\text{d}^3x}{\int \sqrt{-g}T^{\mu\nu}\text{d}^3x},
\end{equation}
where the superscripts $\mu$ and $\nu$ denote the components of the energy-momentum tensor used to define the vortex light' center. Here, $g=\det (g_{\mu\nu})$ represents the determinant of the metric tensor $g_{\mu\nu}$.

Under the weak field approximation, where $h_{\mu\nu} \ll 1$, the gravitational influence on the evolution of vortex light is captured by the perturbation term $\widetilde{A}^\rho$, which is first-order in $h_{\mu\nu}$. Consequently, the energy-momentum tensor for vortex light can be approximated as $T^{\mu\nu} \simeq \bar{T}^{\mu\nu} + \widetilde{T}^{\mu\nu}$. Here, $\bar{T}^{\mu\nu}$, the zeroth-order term in $h_{\mu\nu}$, represents the free motion of vortex light in the absence of gravitational effects. In contrast, $\widetilde{T}^{\mu\nu}$, the first-order term in $h_{\mu\nu}$, accounts for the gravitational influence on the motion of the vortex light. In this case, the leading term in Eq. \eqref{center}, $\int x^i\bar{T}^{\mu\nu}\text{d}^3 x/\int \bar{T}^{\mu\nu}\text{d}^3x$, is independent of gravity and solely represents the free motion of vortex light in the absence of gravitational fields. Our study primarily examines the gravitational effects on the motion of vortex light, leading to a simplified expression for the center of vortex light as:
\begin{equation}
	\langle x^i \rangle^{\mu\nu} \simeq \frac{\int x^i \widetilde{T}^{\mu\nu} \text{d}^3x}{\int\bar{T}^{\mu\nu} \text{d}^3x},
\end{equation}
where the center $\langle x^i\rangle^{\mu\nu}$, defined by $\widetilde{T}^{\mu\nu}$, represents the trajectory deviations in the gravitational field compared to the absence of gravity.

Previous studies have indicated that the choice of components and formulations of the energy-momentum tensor significantly influences the position of the electromagnetic wave packet's center in gravitational fields, affecting the gravitational SHE for light as well \cite{lian2022birefringence}. These findings suggest that the gravitational OHE might be similarly affected by the choice of the energy-momentum tensor components and formulations. To compare with the gravitational SHE, this study investigates the gravitational OHE by analyzing the energy density and energy flux using both symmetric and canonical expressions of the energy-momentum tensor. The respective expressions are defined as follows:
\begin{align}
	&\langle x^i\rangle _{se}\simeq \frac{\int x^i\widetilde{T}^{00}_s \text{d}^3x}{\int \bar{T}^{00}_s \text{d}^3x},\quad \langle x^i\rangle _{sk}\simeq \frac{\int x^i\widetilde{T}^{30}_s \text{d}^3x}{\int \bar{T}^{30}_s \text{d}^3x},\nonumber\\
	& \langle x^i\rangle _{ce}\simeq \frac{\int x^i\widetilde{T}^{00}_c \text{d}^3x}{\int \bar{T}^{00}_s \text{d}^3x},\quad \langle x^i\rangle _{ck}\simeq \frac{\int x^i\widetilde{T}^{30}_c \text{d}^3x}{\int \bar{T}^{30}_c \text{d}^3x},
\end{align}
where $\langle x^i\rangle_{se}$ and $\langle x^i\rangle_{sk}$ respectively denote the centers of energy density and energy flux along the $z$-axis for the symmetric energy-momentum tensor. Similarly, $\langle x^i\rangle_{ce}$ and $\langle x^i\rangle_{ck}$ represent these centers for the canonical energy-momentum tensor.

\subsection{Motion of vortex light with different intrinsic orbital angular momentum in the regime $t\ll b$}

\begin{figure*}   
	\centering 
	\subfigure{\includegraphics[width=0.48\textwidth]{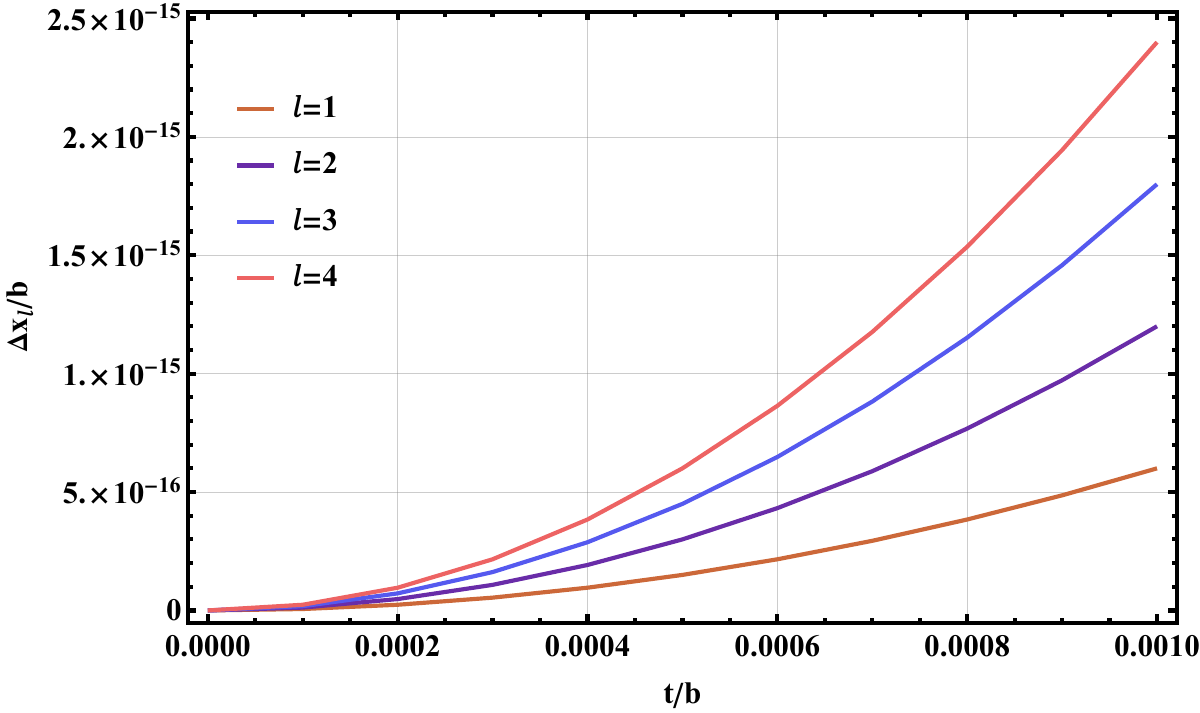}}
	\subfigure{\includegraphics[width=0.48\textwidth]{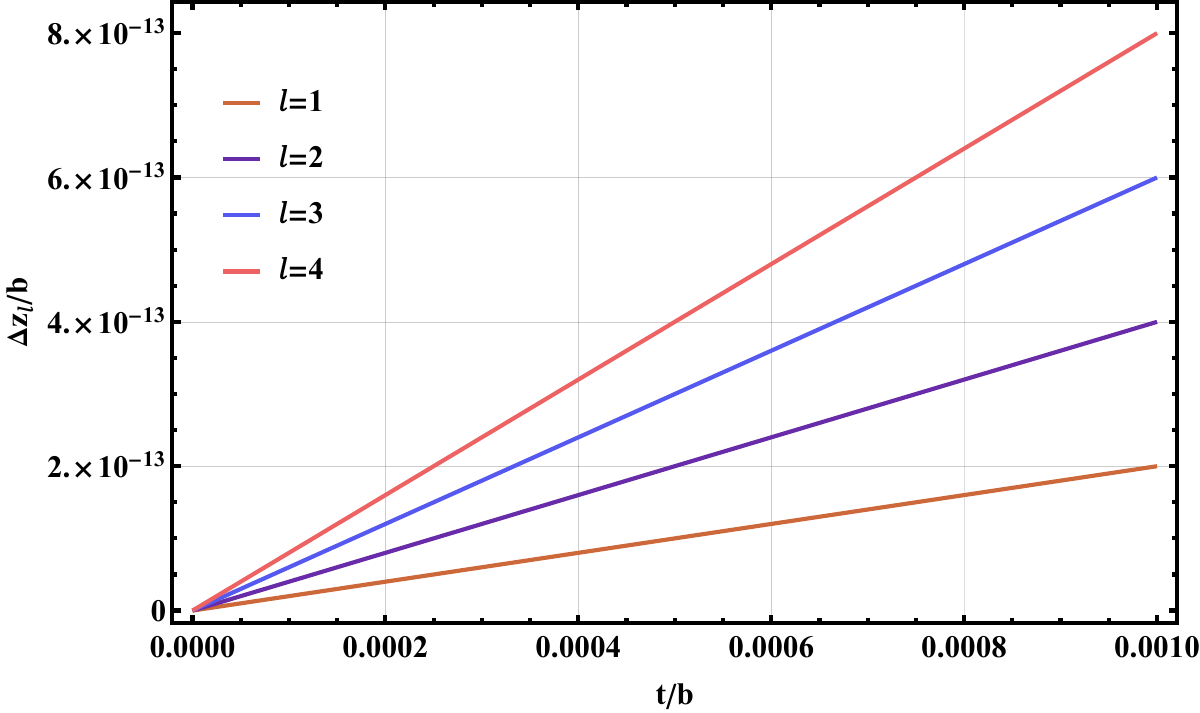}}\\
	\subfigure{\includegraphics[width=0.48\textwidth]{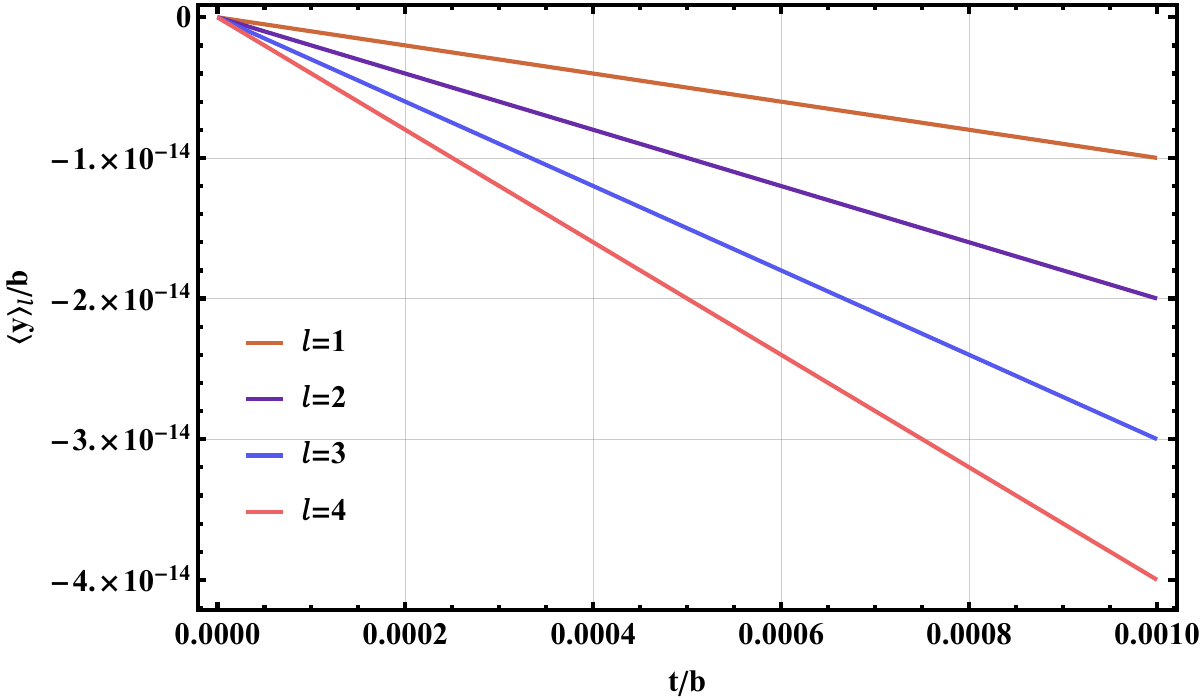}}
	\subfigure{\includegraphics[width=0.48\textwidth]{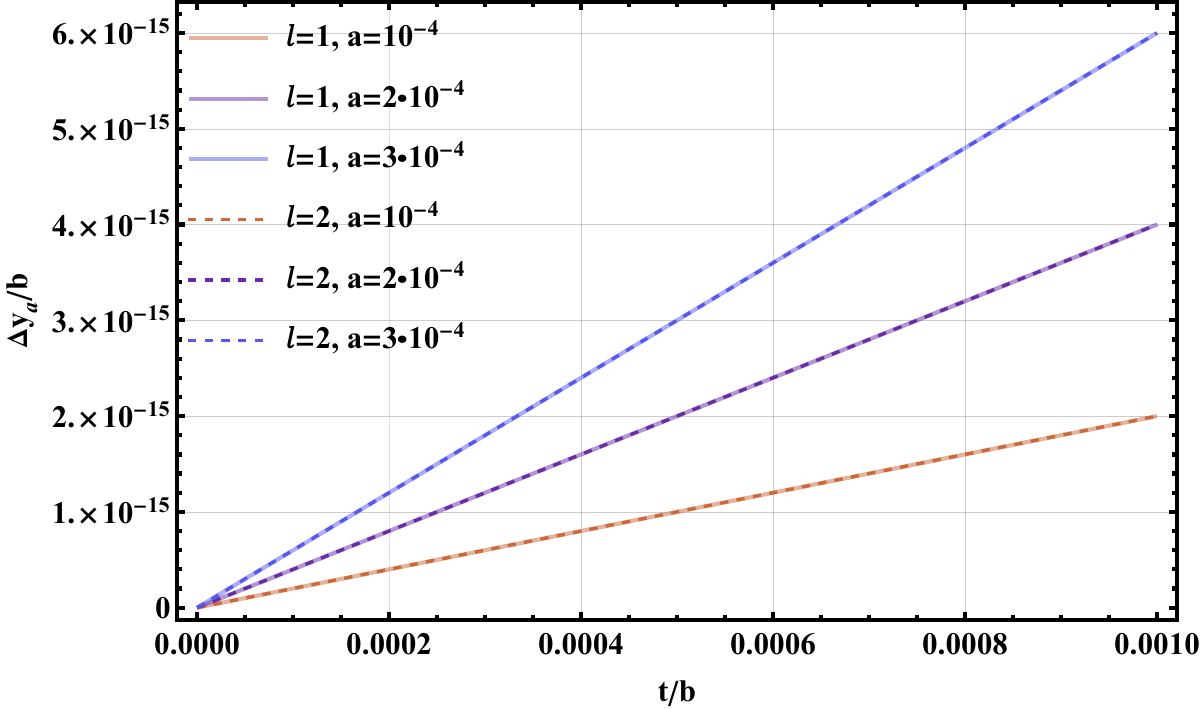}}\\
	\caption{Trajectories of vortex electromagnetic wave packets with intrinsic orbital angular momentum $\ell$ given by using energy density of the symmetric energy-momentum tensor. The upper panels illustrate the separations of the vortex electromagnetic wave packets within the null geodesic plane, defined as $\Delta x_\ell=\langle x\rangle_\ell-\langle x\rangle_{\ell=0}$ and $\Delta z_\ell=\langle z\rangle_\ell-\langle z\rangle_{\ell=0}$. The lower left panel presents the trajectories of the vortex electromagnetic wave packets in the transverse direction, perpendicular to the null geodesic plane. The lower right panel shows the transverse trajectory deviations between rotating ($a\neq 0$) and non-rotating ($a=0$) scenarios, denoted by $\Delta y_a=\langle y\rangle_a-\langle y\rangle_{a=0}$. In these illustrations, parameters are nondimensionalized using $b$ and specified as: $GM/b=10^{-4}$, $p_z b=10^7$, and $\sigma_x b=\sigma_y b=\sigma_z b=5\times 10^3$.}\label{fig.res}
\end{figure*}

Figure \ref{fig.res} shows that the gravitational source's rotation affects the transverse light trajectories without depending on the vortex light's intrinsic orbital angular momentum. Thus, the rotation does not significantly alter the trajectory deviations from the null geodesic for vortex light, indicating a limited sensitivity of the gravitational OHE to the source's rotation. Given this insensitivity, the discussion will not extensively cover the impact of rotation but will instead concentrate on investigating the differences between the gravitational SHE and gravitational OHE.

Unlike the gravitational SHE, where electromagnetic wave packets with different spins primarily diverge along the transverse direction perpendicular to the null geodesic plane, vortex electromagnetic wave packets demonstrate separations both transversely and within the null geodesic plane itself. This distinction highlights a fundamental difference between the gravitational OHE and the gravitational SHE, with the former affecting motion in a broader context.

To compare these two gravitational Hall effects in detail, we examine the transverse motion of vortex electromagnetic wave packets. Specifically, for the vortex electromagnetic wave packet with a travel time $t$ significantly shorter than its initial distance $b$ from the gravitational source ($t \ll b$), the numerical results of its transverse trajectory $\langle y\rangle$, as shown in Fig. \ref{fig.res}, can be fitted numerically by:
\begin{equation}\label{y-l}
	\langle y\rangle_\ell \simeq -\frac{GM\lambda\ell }{2\pi b^2}t,
\end{equation}
where $\lambda = 2\pi / p_z$ represents the wavelength of the electromagnetic wave packet. 

\begin{figure}[htbp]
	\centering 
	\includegraphics[width=0.47\textwidth]{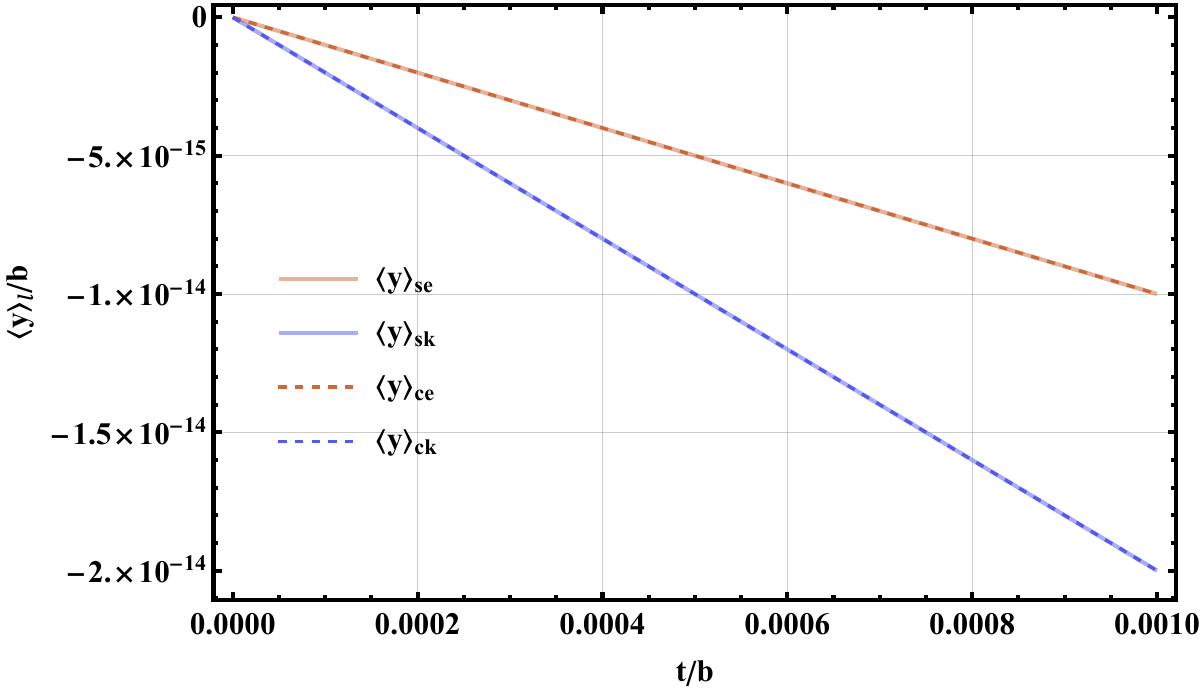}
	\caption{Trajectories of vortex electromagnetic wave packets with intrinsic orbital angular momentum $\ell$ given by using different definitions. Parameters are nondimensionalized using $b$ and specified as: $GM/b=10^{-4}$, $p_z b=10^7$, and $\sigma_x b=\sigma_y b=\sigma_z b=5\times 10^3$. Notations $\langle x^i\rangle_{se}$ and $\langle x^i\rangle_{sk}$ denote the centers of energy density and energy flux along the $z$-axis for the symmetric energy-momentum tensor, while $\langle x^i\rangle_{ce}$ and $\langle x^i\rangle_{ck}$ represent these centers for the canonical energy-momentum tensor.}\label{fig.emt}
\end{figure}

As shown in Fig. \ref{fig.emt}, the gravitational OHE is markedly affected by the choice of energy-momentum tensor components, yet it does not depend on the specific form of the energy-momentum tensor. Therefore, Eq. \eqref{y-l} can be generalized as:
\begin{equation}
	\langle y\rangle_\ell \simeq \frac{\beta GM\lambda\ell }{2\pi b^2}t,
\end{equation}
where $\beta$ reflects the influence of the chosen energy-momentum tensor components on the wave packet's center, with its values provided in Table \ref{tab1}.

\begin{table*}[htbp]
	\centering
	\setlength{\tabcolsep}{2mm}{
	\begin{tabular}{|c|c|c|}
		\hline
		Definitions of the center&$\alpha$&$\beta$\\
		\hline
		Energy density of symmetric energy-momentum tensor& 2.0&-1.0\\
		\hline
		Energy flux density of symmetric energy-momentum tensor& 1.0&-2.0\\
		\hline
		Energy density of  canonical energy-momentum tensor& 1.5&-1.0\\
		\hline
		Energy flux density of canonical energy-momentum tensor& 0.5&-2.0\\
		\hline
	\end{tabular}}
	\caption{Values of $\alpha$ and $\beta$ for the vortex Laguerre-Gaussian electromagnetic wave packet. The values of $\alpha$ have been reported in our previous work \cite{lian2022birefringence}.}\label{tab1}
	\end{table*}

In contrast, the gravitational SHE influences the transverse trajectory as described by Lian et al. \cite{lian2022birefringence}:
\begin{equation}\label{y-s}
	\langle y\rangle_\sigma \simeq \frac{\alpha GM\lambda\sigma}{2\pi b^2}t.
\end{equation}
where $\alpha$ depends on both the selected components and the form of the energy-momentum tensor used to define the wave packet's center, with specific values listed in Table \ref{tab1}. Consequently, the transverse trajectory of an electromagnetic wave packet, carrying a total angular momentum $\sigma+\ell$, can be generalized as follows:
\begin{equation}\label{y-sl}
\langle y\rangle_{\sigma,\ell} \simeq \frac{(\alpha\sigma + \beta\ell) GM\lambda}{2\pi b^2}t.
\end{equation}

Eq. \eqref{y-sl} underlines the difference between the gravitational OHE and gravitational SHE. Specifically, when the center of wave packets is determined using the energy density of symmetric energy-momentum tensor, the coefficients are fitted as $\beta=-1.0$ for gravitational OHE and $\alpha=2.0$ for gravitational SHE. The difference between coefficients $\alpha$ and $\beta$ indicates that the gravitational OHE results in a transverse movement given by $-GM\ell\lambda t/b^2$, whereas the gravitational SHE leads to a movement of $2GM\sigma\lambda t/b^2$. Therefore, an electromagnetic wave packet with a spin of $\sigma$ can be expected to separate from one carrying an intrinsic orbital angular momentum of $\ell$.

Furthermore, the results presented in Table \ref{tab1} demonstrate that both the gravitational SHE and OHE are influenced by the choice of energy-momentum tensor components, suggesting a dependence of these phenomena on the detection method. While the gravitational SHE is sensitive to the specific expressions of the energy-momentum tensor, the gravitational OHE does not exhibit such sensitivity. These observations imply that when coupled to gravity, the intrinsic orbital angular momentum of light may not be directly equivalent to their spin, highlighting the complexity of matter-gravity coupling.

\subsection{The deviation angle of vortex light from the null geodesic plane}

The transverse velocity of the vortex electromagnetic wave packet, which carries a total angular momentum of $\sigma+\ell$, can be derived from Eq. \eqref{y-sl} as:
\begin{equation}
	\frac{d\langle y\rangle_{\sigma,\ell}}{dt}\simeq \frac{(\alpha\sigma + \beta\ell) GM\lambda}{2\pi b^2}.
\end{equation}
 This expression indicates that the wave packet's transverse velocity can be modulated by both its spin $\sigma$ and intrinsic orbital angular momentum $\ell$. Importantly, near the gravitational source, with $t\ll b$, the transverse velocity of the electromagnetic wave packet does not significantly depend on the time $t$. This observation implies that the wave packet's transverse velocity is primarily governed by its distance from the gravitational source.

Consider a vortex electromagnetic wave packet with a total angular momentum of $\sigma+\ell$, emitted from the coordinate origin. At $t=0$, it has an initial transverse velocity of $\text{d}\langle y\rangle_{\sigma,\ell}/\text{d}t=0$. As this wave packet moves away from the gravitational source, predominantly along the $z$-axis within the weak gravitational field, its distance from the gravitational source can be approximately described by $r=\sqrt{b^2+t^2}$. In this case, the wave packet's transverse velocity can be approximated as:
 \begin{equation}
	 \frac{d\langle y\rangle_{\sigma,\ell}}{dt} \sim \frac{(\alpha\sigma + \beta\ell) GM\lambda}{2\pi (b^2+t^2)} - \frac{(\alpha\sigma + \beta\ell) GM\lambda}{2\pi b^2},
 \end{equation}
with higher-order terms than $GM/b$ omitted for simplicity. 

Accordingly, as this wave packet moves away from the gravitational source towards infinity, its transverse velocity might not vanish and can be approximated by:
\begin{equation}
	\frac{\text{d}\langle y\rangle_{\sigma,\ell}}{\text{d}t}\sim - \frac{(\alpha\sigma + \beta\ell) GM\lambda}{2\pi b^2}.
\end{equation}
This persistent transverse velocity suggests a potential angle between the wave packet's trajectory and the null-geodesic plane. This angle, $\theta_{\sigma,\ell}\sim \text{d}\langle y\rangle_{\sigma,\ell}/\text{d}t$, as $t\to +\infty$, can be approximated by:
\begin{equation}
\theta_{\sigma,\ell}\sim - \frac{(\alpha\sigma + \beta\ell) GM\lambda}{2\pi b^2}.
\end{equation}

This equation indicates that light carrying an equal total angular momentum, $\sigma+\ell$, might also follow differing trajectories. Consider three electromagnetic wave packets, each with a total angular momentum of $\sigma+\ell=1$. The first wave packet has a spin $\sigma=1$ and an intrinsic orbital angular momentum $\ell=0$; the second one has $\sigma=0$ and $\ell=1$; and the third has $\sigma=-1$ and $\ell=2$. The angles $\theta_{\sigma,\ell}$ corresponding to these three wave packets are given by:
\begin{equation}\label{exam}
	\theta_{1,0}\simeq -\frac{GM\lambda }{\pi b^2},\quad \theta_{0,1}\simeq \frac{GM\lambda }{2\pi b^2}, \quad \theta_{-1,2}\simeq \frac{2GM\lambda}{\pi b^2},
\end{equation}
where the centers of the three wave packets are defined by using the energy density of the symmetric energy-momentum tensor. Hence, the propagation of light in curved spacetimes is affected not only by its total angular momentum but also by the specific contributions of its spin and intrinsic orbital angular momentum. 

\subsection{Spin vs. intrinsic orbital angular momentum}

The dynamics of light are governed by Maxwell's equations and can be described through electromagnetic wave packets. For a spin-polarized electromagnetic wave packet, its spin is characterized by the polarization of the electromagnetic field $A^{\mu}$, specifically represented by a polarization vector $\vec{\epsilon}$. Importantly, $\vec{\epsilon}$ could be independent of the wave packet's spatial distribution, as suggested by left-handed and right-handed circularly polarized light beams. This independence suggests that its spin remains physically meaningful when the wave packet is approximated as a point-like particle.

However, for vortex light, its intrinsic orbital angular momentum is captured by a phase factor $\psi (\vec{x},t) \propto \exp (i\ell \phi)$ within its electromagnetic wave packet, where $\phi$ is the azimuthal angle. This angle, and consequently the phase factor, are dependent on the wave packet's spatial distribution. Thus, when the wave packet's size is theoretically reduced to zero, the phase factor becomes undefined, suggesting that the intrinsic orbital angular momentum might lose its physical meaning in the approximation of light as point-like particles.

The MPD equations are widely employed to analyze the dynamics of particles with angular momentum, particularly for describing the motion of spin-polarized light. In these equations, the characteristics of light are simplified to angular momentum, employing a geodesic-like equation for motion description. However, by reducing dynamics of electromagnetic wave packets to geodesic-like equations, this method simplifies the spatial distribution of these wave packets into a mere representation of angular momentum. Consequently, it fails to adequately capture the intrinsic orbital angular momentum of vortex light, which is significantly influenced by the spatial distribution of its wave packets. This discussion suggests that the MPD equations may have limitations in accurately describing the motion of vortex light, possessing intrinsic orbital angular momentum. Our investigation into the gravitational OHE for vortex light highlights these limitations.

\section{Discussion}\label{sec.dis}

In this study, we explore the dynamics of vortex light, modeled as vortex Laguerre-Gaussian electromagnetic wave packets, within the Lense-Thirring metric by solving Maxwell's equations. We determine the trajectory of these wave packets by analyzing the center of their energy-momentum tensor. Unlike the gravitational SHE, where light with opposite spins tends to diverge perpendicular to the null geodesic plane, vortex light exhibiting different intrinsic orbital angular momentum also demonstrates separation within this plane. This separation, driven by the intrinsic orbital angular momentum, is identified as the gravitational OHE. This distinction between the gravitational spin and orbital Hall effects suggests that, in gravitational fields, the intrinsic orbital angular momentum of light may not align with their spin, highlighting a complex coupling between angular momentum of light and gravity.

For spin-polarized vortex light originating a finite distance $r_0$ away from a gravitational source and propagating to infinity, its deviation angle $\theta_{\sigma,\ell}$ from the null geodesic plane can be effectively approximated by the expression $\theta_{\sigma,\ell} \sim -(\alpha\sigma + \beta\ell)GM\lambda / 2\pi r_0^2$. The symbols $\sigma$ and $\ell$ represent the spin and intrinsic orbital angular momentum of the vortex light, respectively. The coefficients $\alpha$ and $\beta$ depend on the chosen expression and component of the energy-momentum tensor that are used to define the center of the light's wave packet. Specifically, when the energy density of the symmetric energy-momentum tensor is applied, these coefficients are established as $\alpha = 2.0$ and $\beta = -1.0$. This leads to a specific expression for the deviation angle: $\theta_{\sigma,\ell} \sim -(2.0\sigma - 1.0\ell)GM\lambda / 2\pi r_0^2$.

The findings reveal that light, even when characterized by an equivalent total angular momentum $\sigma+\ell=0$, can exhibit diverse trajectories upon propagating through a gravitational field. In detail, an electromagnetic wave packet with a spin $\sigma=0$ and and an intrinsic orbital angular momentum $\ell=0$ is anticipated to adhere to the null geodesic, yielding a zero deviation angle $\theta_{0,0}=0$. In contrast, a wave packet with spin $\sigma=1$ and intrinsic orbital angular momentum $\ell=-1$ is predicted to diverge from this path, displaying a deviation angle $\theta_{1,-1}\sim -3GM\lambda/2\pi r_0^2$. Hence, despite sharing the same total angular momentum, the propagation paths of light can vary, indicating a non-zero separation angle.

The intrinsic orbital angular momentum of vortex light is characterized as a phase factor $\psi\propto \exp (i\ell\phi)$ in its electromagnetic wave packet, with $\ell\phi$ indicating the dependence on the spatial distribution of its wave packet. Our findings indicate that, when coupled to gravity, the intrinsic orbital angular momentum of light might not align with its spin. This discrepancy suggests that the spatial distribution of the vortex light extends beyond a simple angular momentum representation, necessitating a more detailed examination of its wave packet characteristics.

Tamburini et al. have demonstrated that light can manifest in vortex states with intrinsic orbital angular momentum when emitted from the vicinity of a rotating black hole \cite{Tamburini:2011tk}. Further investigations have shown that the trajectory of such vortex light can be influenced by the rotation of the black hole, as reported in subsequent studies \cite{Tamburini:2019vrf,PhysRevA.104.013718,andersson2023spin}. Consequently, in the strong gravitational fields surrounding compact stars and black holes, the gravitational OHE is expected to be modulated by the rotation of the gravitational source. This phenomenon could potentially offer a novel approach for observing the rotation of compact stars and black holes.

\section*{Acknowledgement}\label{Acknowledgement}
We are grateful to Bei Liu, Li-Li Yang and Qi-Liang Zhao for the helpful discussions, and we thank Peng-Cheng Zhao for his suggestions during the numerical calculations. This work was supported by the Fundamental Research Funds for the Central Universities, Sun Yat-sen University.

\section*{Appendix: Equations of motion for vortex light}\label{sec.app1}

Within the weak field approximation, characterized by $GM/r \ll 1$, the dynamics of the perturbation term $\widetilde{A}^\nu (\vec{x},t)$ for vortex light are governed by Eq. \eqref{eom-first}. By adhering to the first order of $x^i/b$ as stipulated by Eqs. \eqref{h-beg}-\eqref{h-end}, Eq. \eqref{eom-first} simplifies to
\begin{strip}
\begin{align}\label{eom-spe}
		\left(-\partial_x^{2} +\nabla^{2}\right) \tilde{A}^{0}&=\frac{2 G M}{b^{3}}\left(z \partial_{t} \bar{A}^3+(b-2 x) \partial_{t} \bar{A}^1+y \partial_{t} \bar{A}^{2}\right)+\frac{2 G Ma}{b^{5}}(-6 b y(\partial_{x} \bar{A}^1-\partial_{y} \bar{A}^{2}))\nonumber\\
		&+\frac{2 G Ma}{b^{5}}(3 (b^{2}-3 b x)(\partial_{y} \bar{A}^{1}+\partial_{x} \bar{A}^{2})+3 b z(\partial_{z} \bar{A}^{2}+\partial_{y} \bar{A}^{3})),\\
		\left(-\partial^{2}_ t+\nabla^{2}\right) \tilde{A}^{1}&=\frac{2G M}{b}\left(\partial^2_{t} \bar{A}^1+\partial_i\partial^i \bar{A}^1\right)+\frac{2 G M}{b^{3}}(2(b-2 x) \partial_{x} \bar{A}^1+y\left(\partial_{y} \bar{A}^ 1+\partial _x \bar{A}^2\right))\nonumber\\
		 &+\frac{2 G M z}{b^{3}}\left(\partial_{z} \bar{A}^ 1+ \partial_ x \bar{A}^3\right)+ 
		\frac{4 a G M}{b^{3}}\left(-y \partial_{t, x} \bar{A}^ 1+(b-2 x) \partial_{t, y} \bar{A}^ 1\right)\nonumber\\
		&+\frac{2 a G M}{b^{3}}\left(1-\frac{3 a x}{b}\right) \partial_{t} \bar{A} ^2,\\
		\left(-\partial_{t}^{2}+\nabla^{2}\right) \tilde{A}^{2}&=\frac{2 G M}{b}\left(\partial_{t}^2 \bar{A}^{2}+\partial_i\partial^i \bar{A}^{2}\right)+ 
		 \frac{2 G M}{b^{3}}(2 y \partial_{y} \bar{A}^{2}+(b-2 x)(\partial_{x} \bar{A}^{2}+\partial_{y} \bar{A}^{1}))\nonumber\\
		 & +\frac{2 G M z}{b^{3}}(\partial_{z} \bar{A}^{2}+\partial_{y} \bar{A}^{3})+\frac{2 G M a}{b^{3}}((1-\frac{3 x}{b}) \partial_{t} \bar{A}^1+ \frac{3 z}{b} \partial_{t} \bar{A}^{3})\nonumber\\
		 &+\frac{4 G M a}{b^{3}}\left(-y \partial_{t,x} \bar{A}^{2}+(b-2 x) \partial_{t, y} \bar{A}^{2}\right),\\
		(-\partial_{t}^{2}+\nabla^{2}) \tilde{A}^{3}& =\frac{2 G M}{b}(\partial_{t}^{2} \bar{A}^{3}+\partial_i\partial^i\bar{A}^{3})+\frac{2 G M}{b^{3}}(2 z \partial_{z} \bar{A}^{3}+(b-2 x)(\partial_ y \bar{A} ^{1}+\partial_{x} \bar{A}^{2}))\nonumber\\
		& +\frac{2 G M y}{b^{3}}(\partial_{z} \bar{A}^{2}+\partial_{y} \bar{A}^{3})+\frac{6 G M a}{b^{5}}(y \partial_{t} \bar{A}^{1}- (b-4 x) \partial_{t} \bar{A}^{2})\nonumber\\
		&+\frac{4 G M a}{b^{3}}(-y \partial_{t, x} \bar{A}^{3}+(b-2 x) \partial_{t, y} \bar{A}^{3}), \label{eom-sped}
\end{align}
\end{strip}
where $i=1,2,3$ denotes the spatial indices. 

Given the complexity of solving second-order inhomogeneous partial differential equations, especially due to their challenging nature even with numerical methods, we opt for a three-dimensional spatial Fourier transform to facilitate their analysis. Hence, Eqs. \eqref{eom-spe}-\eqref{eom-sped} become to

\begin{strip}
\begin{align}\label{Equations of motion in Cases I(Rotating axis along z direction)2}
		\left(\partial_{t}^{2}+\omega^{2}\right) \widetilde{A}_{f}^{0}
		&=\frac{2 G M}{b^{2}}\left(\left(1-\frac{2 i}{b} \frac{\partial}{\partial k_{x}}\right)\left(i \omega \bar{A}_{f}^{1}\right)+\frac{i}{b} \frac{\partial}{\partial k_{y}}\left(i \omega \bar{A}_{f}^{2}\right)+\frac{i}{b} \frac{\partial}{\partial k_{z}}\left(i \omega \bar{A}_{f}^{3}\right)\right)\nonumber\\
		&+\frac{6 G M a}{b^{3}}\left(\frac { i } { b } \frac { \partial } { \partial k _ { y } } (i k_{x} \bar{A}_{f}^{1}-i k_{y} \bar{A}_{f}^{2})-\left(1-\frac{3 i}{b} \frac{\partial}{\partial k_{x}}\right)(i k_{x} \bar{A}_{f}^{2}+i k_{y} \bar{A}_{f}^{1})\right)\nonumber\\
		&-\frac{6 G M a}{b^3}\left(\frac{i}{b} \frac{\partial}{\partial k_{z}}\left(i k_{z} \bar{A}_{f}^{2}+i k_{y} \bar{A}_{f}^{3}\right)\right)\\
		\left(\partial_{t}^{2}+\omega^{2}\right) \widetilde{A}_{f}^{1}
		&=\frac{4 G M}{b}\left(1-\frac{i}{b} \frac{\partial}{\partial k_{x}}\right)(\omega^{2} \bar{A}_{f}^{1})+\frac{2 G M}{b^{2}}\left(\frac{\partial ( k_{z} \bar{A}_{f}^{1}+ k_{x} \bar{A}_{f}^{3})}{b\partial k_{z}}\right)-\frac{4i G M}{b^{2}}\nonumber\\
		&\times\left(\left(1-\frac{2 i}{b} \frac{\partial}{\partial k_{x}}\right)( k_{x} \bar{A}_{f}^{1})+ \frac{i\partial ( k_{x} \bar{A}_{f}^{2}+ k_{y} \bar{A}_{f}^{1})}{2b\partial k_{y}}\right)-\frac{2i G M a}{b^{3}}\left(1-\frac{3 i}{b} \frac{\partial}{\partial k_{x}}\right)\nonumber\\
		&\times ( \omega \bar{A}_{f}^{2})+\frac{4 G M a}{b^{3}}\left( \frac{i \partial}{\partial k_{y}}( \omega k_{x} \bar{A}_{f}^{1})\right)-\frac{4 G M a}{b^{2}}\left(1- \frac{2i\partial}{b\partial k_{x}}( \omega k_{y} \bar{A}_{f}^{1})\right)\\
		\left(\partial_{t}^{2}+\omega^{2}\right) \widetilde{A}_{f}^{2}
		&=\frac{4 G M}{b}\left(1-\frac{i}{b} \frac{\partial}{\partial k_{x}}\right)\left(\omega^{2} \bar{A}_{f}^{2}\right)-\frac{2 G M}{b^{2}}\left(\left(1-2 \frac{i}{b} \frac{\partial}{\partial k_{x}}\right)(i k_{y} \bar{A}_{f}^{1}+i k_{x} \bar{A}_{f}^{2})\right)\nonumber\\
		&-\frac{2 G M}{b^{2}}\left(\frac{2\partial (i k_{y} \bar{A}_{f}^{2})}{b\partial k_{y}}-\frac{\partial (k_{z} \bar{A}_{f}^{2}+ k_{y} \bar{A}_{f}^{3})}{b\partial k_{z}}\right) +\frac{2i G M a}{b^{3}}\left(1-\frac{3 i}{b} \frac{\partial}{\partial k_{x}}\right)( \omega \bar{A}_{f}^{1})\nonumber\\
		&-\frac{2 G M a}{b^{4}}\frac{\partial ( \omega \bar{A}_{f}^{3})}{\partial k_{z}}-\frac{4 G M a}{b^{2}}\left(- \frac{i \partial ( \omega k_{x} \bar{A}_{f}^{2})}{\partial k_{y}}+\left(1-\frac{2i}{b} \frac{\partial}{\partial k_{x}}\right)( \omega  k_{y} \bar{A}_{f}^{2})\right) \\
		\left(\partial_{t}^{2}+\omega^{2}\right) \widetilde{A}_{f}^{3}&=\frac{4 G M}{b}\left(1-\frac{i}{b} \frac{\partial}{\partial k_{x}}\right)(\omega^{2} \bar{A}_{f}^{3})-\frac{2i G M}{b^{2}}\left(1-\frac{2 i}{b} \frac{\partial}{\partial k_{x}}\right)( k_{z} \bar{A}_{f}^{1}+ k_{x} \bar{A}_{f}^{3})\nonumber\\&
		+\frac{2 G M}{b^{3}}\left(\frac{\partial ( k_{y} \bar{A}_{f}^{3}+ k_{z} \bar{A}_{f}^{2})}{\partial k_{y}}+\frac{2\partial (k_{z} \bar{A}_{f}^{3})}{\partial k_{z}}\right)+\frac{6 G M a}{b^{4}}\left( \frac{\partial ( \omega \bar{A}_{f}^{2})}{\partial k_{z}}\right)\nonumber\\
		&-\frac{4 G M a}{b^{2}}\left(\left(1-\frac{2 i}{b} \frac{\partial}{\partial k_{x}}\right)( \omega k_{y} \bar{A}_{f}^{3})-\frac{i}{b} \frac{\partial}{\partial k_{y}}(\omega k_{x} \bar{A}_{f}^{3})\right),
\end{align}
\end{strip}
where $\widetilde{A}^\rho_f$ is defined as
\begin{equation}
	\widetilde{A}^\rho_f=\frac{1}{(2\pi)^3}\int\widetilde{A}^\rho(\vec{x},t)\exp(-i\vec{k}\cdot\vec{x})\text{d}^3x.
\end{equation}

\bibliography{reference.bib} 
\end{document}